\documentclass[twocolumn,prl,showpacs,superscriptaddress]{revtex4}

\usepackage{graphicx}
\usepackage{dcolumn}
\usepackage{bm}

\begin{document}

\title{New Trial Wave Functions for Quantum Hall States at Half Filling}
\author{Jian Yang}
\email{jianyang1234@gmail.com}
\altaffiliation{On leave from Schlumbeger, Houston Formation
Evaluation Center, Sugar Land, TX 77478, USA.}
\affiliation{Department of Physics and Center of Theoretical and
Computational Physics, The University of Hong Kong, Pokfulam Road,
Hong Kong, China}

\begin{abstract}

New trial wave functions corresponding to half filling quantum Hall
states are proposed. These wave functions are constructed by first
pairing up the quasielectrons of the $1/3$ Laughlin quantum Hall
state, with the same relative angular momentum for each pair, and
then making the paired quasielectrons condense into a 1/4 Laughlin
state. The quasiparticle excitations of the proposed wave functions
carry $\pm1/4$ of electron charge, and obey Abelian fractional
statistics. In the spherical geometry, the total flux quanta
$N_{\phi}$ is shown to be related to the number of electrons $N$ by
$N_{\phi} = 2N-(5-q)$ with $q$ being the relative angular momentum
between the quasielectrons in each pair which takes values of
non-negative even integers. The overlaps are calculated between the
proposed trial wave functions, including the ground state,
quasiexciton states, and quasihole states, and the exact states of
the finite size systems at $N_{\phi} = 2N-3$. The near unity
overlaps are obtained in the lowest Landau level, while the moderate
overlaps are obtained in the second Landau level. The relevance of
the wave functions to the yet to be discovered fractional quantum
Hall effect in the lowest Landau level, as well as the $5/2$ quantum
Hall effect is also discussed.
\end{abstract}
 \maketitle

The observed fractional quantum Hall effect (FQHE) at
$\nu=5/2$\cite{Willett}, which deviates from the odd denominator
filling factor rule, calls for a new mechanism that is different
from its odd denominator filling factor counterparts. The nature of
the $5/2$ FQHE has been under intensive study recently, in part due
to the exotic non-Abelian statistics of the quasiparticle (QP)
excitations of the Moore-Read state\cite{MR}, which is considered to
be the leading candidate for the ground state. Physically, this new
state can be regarded as a p-wave BCS pairing state, apart from a
Jastrow factor that attaches two flux to each of the
electrons\cite{MR}\cite{Greiter}. It has been shown through finite
size studies that the MR wave functions have only moderate overlaps
with the exact $2/5$ ground state\cite{Morf}, in particular with the
low energy states\cite{Sreejith}, in the second Landau level (SLL),
although the overlaps can be improved by adjusting the
pseudopotentials\cite{Morf} or fine tuning the finite thickness of
the layer that confined the two dimensional (2D)
electrons\cite{Peterson}.

On the other hand, the MR wave function has a rather poor overlap
with the exact state in the lowest Landau level (LLL). While it is
generally believed that a 2D electron system at half filling forms a
compressible Fermi liquid type state\cite{Halperin} in the LLL, it
does not rule out a possibility that under certain conditions when
the short range interactions are softened, an incompressible state
may emerge. In fact, recent studies have shown such a possibility by
including the finite width of the layer confining the 2D
electrons\cite{Peterson}.

The moderate overlaps of the MR states with the exact states in the
SLL, and the poor overlaps with the exact states in the LLL, in
contrast with the fact that the Laughlin wave functions find near
unity overlaps with their corresponding exact states, justify a
continuing pursue of alternatives to the MR states. It is noted that
the MR's particle-hole conjugate state and the superposition of both
have also been studied recently\cite{Levin}\cite{Lee}\cite{Wang}.

In this paper, new trial wave functions corresponding to half
filling quantum Hall effect are proposed. These wave functions are
constructed by first pairing up the quasielectrons of the 1/3
Laughlin quantum Hall state, requiring the same relative angular
momentum for each pair, and then making the paired quasielectrons
condense into a 1/4 Laughlin state. Hereafter, we will use the
notation LQE refering to the quasielectrons of the Laughlin state,
and PQE to the paired quasielectrons of the Laughlin state. It is
shown that the proposed wave functions support their own QP
excitations, which carry $\pm1/4$ of electron charge, and obey
$\pm\pi/4$ Abelian fractional statistics. The new wave functions are
explicitly constructed in the spherical geometry, with the total
number of flux quanta $N_{\phi}$ and the number of electrons $N$
related by $N_{\phi}=2N-(5-q)$, where $q$ is the relative angular
momentum between the LQEs in each pair, and takes values of zero and
positive even integers. The wave functions of the low energy states
at half filling (quasiexciton states) as well as away from half
filling (quasiparticle states), are also constructed explicitly.
Their overlaps with the exact states of finite systems with
$N_{\phi}=2N-3$ using the exact diagonalization techniques are
calculated. The near unity overlaps in the lowest Landau level (LLL)
are obtained. Their overlaps with the exact states in the SLL are
found to be moderate. This is in contrast with the MR wave function,
where it finds a larger overlap with the exact state in the SLL over
the LLL.

It should be pointed out that the idea of pairing up the LQEs and
forming their own Laughlin state at half filling was proposed before
by the author, Su and Su\cite{YangSuSu}, and also discussed by
others\cite{Wen}. However, the exact mechanism to pair up the LQEs,
the dependence of the $N_{\phi}$-$N$ relationship on the relative
angular momentum of the LQEs in each pair, the explicit construction
of the ground state wave functions, the quasiexciton wave functions,
and the $1/4$ charged quasiparticle wave functions, as well as their
overlaps with the exact states of the finite size systems are all
new contributions in this paper.

In Haldane's spherical geometry, the Laughlin wave function can be
written as\cite{Haldane}:
\begin{equation}
\label{Laughlin} \Phi_{m_L}={\prod}_{i<j}(u_iv_j-v_iu_j)^{m_L},
\end{equation}
where $(u, v)$ are the spinor variables describing electron
coordinates, and $m_L$ is an odd integer as the result of the Pauli
exclusion principle. The total flux quanta of the system $N_{\phi}$
described by the Laughlin wave function in Eq.(\ref{Laughlin}) is
$N_{\phi} = m_L(N-1)$. When $N_{\phi}$ is decreased by an amount of
$N_{qe}$, there are $N_{qe}$ LQEs generated, described
by\cite{Laughlin}\cite{Haldane}:
\begin{equation}
\label{LQE_WF}
S^{qe}({\alpha}_1,{\beta}_1)S^{qe}({\alpha}_2,{\beta}_2){\ldots}S^{qe}({\alpha}_{N_{qe}},{\beta}_{N_{qe}})\Phi_{m_L},
\end{equation}
where the LQE operator is
\begin{equation}
\label{LQE_OP}
S^{qe}({\alpha},{\beta})={\prod}_{j=1}^{N}({\beta}^*\frac{\partial}{{\partial}u_j}-{\alpha}^*\frac{\partial}{{\partial}v_j}),
\end{equation}
with $({\alpha},{\beta})$ being the spinor variables describing the
LEQ coordinates. It has been shown numerically that the LQE wave
functions Eq.(\ref{LQE_WF}) provide a rather exact description for
the low energy states from a single LQE to multiple LQEs, and all
the way to the $2/5$ filling factor, at which a new hierarchy state
is formed when the LQEs form their own Laughlin
state\cite{He}\cite{YangHierarchy}. This is in part due to the fact
that a LQE behaves the same as a charged particle in its own LLL
with the total flux quanta equal to the total number of the
underlying electrons $N$. This can be best seen by expanding
$S^{qe}({\alpha},{\beta})$ in terms of $({\alpha},{\beta})$:
\begin{equation}
\label{LQE_TF}
S^{qe}({\alpha},{\beta})={\sum}_{m=-\frac{N}{2}}^{\frac{N}{2}}{\psi}_{\frac{N}{2},m}({\alpha},{\beta})G_{\frac{N}{2},m}^{qe},
\end{equation}
where ${\psi}_{\frac{N}{2},m}({\alpha},{\beta})$ is the LLL wave
function of a particle which sees the total flux quanta $N$, and
$G_{\frac{N}{2},m}^{qe}$ is of the form
\begin{eqnarray}
&&G_{\frac{N}{2},m}^{qe}=(-1)^{\frac{N}{2}-m}[\frac{N!}{(\frac{N}{2}+m)!(\frac{N}{2}-m)!}]^{-1/2}\cdot
 \nonumber \\
&&
\sum_{1{\leq}l_1<l_2<{\ldots}{\leq}l_{\frac{N}{2}+m}}\frac{\partial}{\partial
v_{l_1}}\frac{\partial}{\partial
v_{l_2}}\ldots\frac{\partial}{\partial v_{l_{\frac{N}{2}+m}}} \cdot\nonumber \\
&& \prod_{l({\neq} l_1, l_2, \ldots, l_{\frac{N}{2}+m})}
\frac{\partial}{\partial u_l}.
\end{eqnarray}
When applied to the Laughlin wave function, $G_{\frac{N}{2},m}^{qe}$
will generate a LQE with angular momentum $(L,L_z)=(\frac{N}{2},m)$.

Now we come to the main part of this paper. Suppose we have an even
number of LQEs generated:
\begin{equation}
(G_{\frac{N}{2},m_1}^{qe}G_{\frac{N}{2},m_2}^{qe})(G_{\frac{N}{2},m_3}^{qe}G_{\frac{N}{2},m_4}^{qe})...(G_{\frac{N}{2},m_{N_{qe}-1}}^{qe}G_{\frac{N}{2},m_{N_{qe}}}^{qe})\Phi_{m_L}.
\end{equation}
In the above equation, we have bracketed the
$G_{\frac{N}{2},m}^{qe}$s in pair, as we would like to pair up $m_1$
with $m_2$, $m_3$ with $m_4$, ..., and $m_{N_{qe}-1}$ with
$m_{N_{qe}}$. Since the angular momentum for each LQE is $N/2$, the
total angular momentum for each pair is therefore $L_p=N-q$ with $q$
being a non-negative even integer representing a relative angular
momentum of the LQE pair. The following wave function:
\begin{equation}
G_{L_p,p_1}^{pqe}G_{L_p,p_2}^{pqe}{\ldots}G_{L_p,p_{N_{p}}}^{pqe}\Phi_{m_L}.
\end{equation}
where
\begin{equation}
G_{L_p,p}^{pqe}={\sum}_{m_1,m_2}<L_p,p|\frac{N}{2}m_1;\frac{N}{2}m_2>G_{\frac{N}{2},m_1}^{qe}G_{\frac{N}{2},m_2}^{qe},
\end{equation}
and $<L_p,p|\frac{N}{2}m_1;\frac{N}{2}m_2>$ being Clebsch-Gordon
coefficients, describe $N_p=N_{qe}/2$ PQEs, each PQE has angular
momentum $L_p=N-q$ with it's $z$ component $p$ taking values from
$-L_p$ to $L_p$. Similar to Eq.(\ref{LQE_TF}), one can define a new
operator corresponding to $G_{L_p,p}^{pqe}$ as
\begin{equation}
\label{PQE_TF}
S^{pqe}({\xi},{\eta})={\sum}_{p=-L_p}^{L_p}{\psi}_{L_p,p}({\xi},{\eta})G_{L_p,p}^{pqe},
\end{equation}
where ${\psi}_{L_p,p}({\xi},{\eta})$ is the LLL wave function of a
particle which sees the total flux quanta $2L_p=2(N-q)$.

It is the key assumption of this paper that the PQEs described by
$S^{pqe}({\xi},{\eta})$ condense into their own Laughlin state to
form a new state of the system, described by the following wave
functions:
\begin{equation}
\label{GROUND} \int
d\Omega{\prod}_{1{\leq}a<b{\leq}N_p}({\xi}_a{\eta}_b-{\xi}_b{\eta}_a)^{2n}
{\prod}_{a=1}^{N_p}S^{pqe}({\xi}_a,{\eta}_a)\Phi_{m_L}.
\end{equation}
where $d\Omega = {\prod}_{a=1}^{N_p}d{\Omega}_a =
{\prod}_{a=1}^{N_p}\sin(\vartheta_a)d{\vartheta_a}d{\varphi_a}$ if
we let the spinor variables
$({\xi},{\eta})=(\cos(\vartheta/2)e^{i\varphi/2},\sin(\vartheta/2)e^{-i\varphi/2}$,
and $n$ is a positive integer. The filling factor of the electron
system described by Eq.(\ref{GROUND}) can be derived by the the
equation between $N_{\phi}$ and $N_p$
\begin{equation}
N_{\phi}=m_L(N-1)-2N_p,\label{NPHI}
\end{equation}
which describes $2N_p$ LQE created on top of the $1/m_L$ Laughlin
state, and that between the total number of the flux quanta seen by
the PQEs $2(N-q)$ and $N_p$:
\begin{equation}
2(N-q)=2n(N_p-1),\label{NPHIB}
\end{equation}
which describes the PQEs condense into their own $1/(2n)$ Laughlin
state. Combining these two equations and setting $m_L=3$ and $n=2$,
we arrive at:
\begin{equation}
\label{NPHI_N}
N_{\phi} = 2N-(5-q),
\end{equation}
which gives $1/2$ filling factor in the thermodynamic limit. The
"shift" defined as $S=2N-N_{\phi}$\cite{Wen2} is given by $5-q$,
which can take values $5$, $3$, $1$, etc., corresponding to the
relative angular momentum $0$, $2$, $4$, etc. The case of $S=3$ is
of particular interest for two reasons. First of all, it is widely
believed to be responsible for the $5/2$ FQHE observed
experimentally\cite{Morf}. Secondly, from the potential energy point
of view, it seems more favorable to have a non-zero relative angular
momentum between the LEQs in each pair to avoid them on top of each
other (which corresponds to the zero relative angular momentum,
i.e., the shift $S=5$). On the other hand, the relative angular
momentum which corresponds to the separation of the LEQs in each
pair shall not be too large in order to justify their condensation
into a Laughlin state. The relative angular momentum $q=2$ which
corresponds to the shift $S=3$ is therefore seems to be an optimal
choice, although there is no solid reason to rule out other choices
of the relative angular momentum.

The charge of the QPs supported by the wave functions
Eq.(\ref{GROUND}) can be calculated by starting with Eq.(\ref{NPHI})
and increasing (decreasing) the number of electrons by 2 in order to
keep all the LQEs paired up, and examining the total number of the
shortage (excess) of the flux quanta seen by the PQEs in
Eq.(\ref{NPHIB}), which turns out to be $8$. This corresponds to
$\pm 1/4$ of the electron charge. The statistics is the Abelian
fractional statistics $\pm\pi/4$.

It is well known that the Laughlin state supports a quasiexciton as
its low energy excitations. By the same reasoning we propose the
following quasiexciton wave functions to describe the low energy
excitations at half filling:
\begin{eqnarray}
\label{EXCITON} && \int
d\Omega{s^{qe}({\mu}_1,{\nu}_1)s^{qh}({\mu}_2,{\nu}_2)}{\prod}_{1{\leq}a<b{\leq}N_p}({\xi}_a{\eta}_b-{\xi}_b{\eta}_a)^{2n}
\nonumber \\
&& {\prod}_{a=1}^{N_p}S^{pqe}({\xi}_a,{\eta}_a)\Phi_{m_L}.
\end{eqnarray}
where $s^{qe}({\mu},{\nu})$ is the quasielectron operator of the PQE
Laughlin state, similar to Eq.(\ref{LQE_OP}):
\begin{equation}
s^{qe}({\mu},{\nu})={\prod}_{a=1}^{N_p}({\nu}^*\frac{\partial}{{\partial}\xi_a}-{\mu}^*\frac{\partial}{{\partial}\eta_a}).
\end{equation}
and $s^{qh}({\mu},{\nu})$ is the quasihole operator of the PQE
Laughlin state,
\begin{equation}
s^{qh}({\mu},{\nu})={\prod}_{a=1}^{N_p}({\nu}\xi_a-{\mu}\eta_a).
\end{equation}
where the spinor ($\mu,\nu$) represents the coordinates of the
quasielectron or the quasihole of the PQE Laughlin state. The
angular momentum described by Eq.(\ref{EXCITON}) is from $1$, $2$,
..., to $N_p$.

It can be shown that the minimum number, which is 2, of $1/4$
fractionally charged QHs can be created when $N$ is an odd number
and the flux quanta $N_{\phi}$ is increased by $1$ from
Eq.(\ref{NPHI_N}). The corresponding wave function can be described
by
\begin{eqnarray}
\label{TWOQH} && \int
d\Omega{s^{qh}({\mu}_1,{\nu}_1)s^{qh}({\mu}_2,{\nu}_2)}{\prod}_{1{\leq}a<b{\leq}N_p}({\xi}_a{\eta}_b-{\xi}_b{\eta}_a)^{2n}
\nonumber \\
&& {\prod}_{a=1}^{N_p}S^{pqe}({\xi}_a,{\eta}_a)\Phi_{m_L}.
\end{eqnarray}
In the angular momentum space, it forms independent states with the
total angular momentum equal to $N_p$, $N_p-2$, ..., and $1$.

In order to validate the proposed wave functions, we have calculated
the overlaps of the ground state wave functions Eq.(\ref{GROUND}),
quasiexciton wave functions Eq.(\ref{EXCITON}), and the two QH wave
functions Eq.(\ref{TWOQH}) with the exact states of the finite
systems at $N_{\phi}=2N-3$ using the exact diagonalization technique
in the spherical geometry. In Fig.1(a), we plot the energy spectrum
in an arbitrary units of a $(N_{\phi},N)=(9,6)$ finite system in the
LLL versus angular momentum $L$. The numbers ($0.9980$ and $0.9676$)
on top of the two energy bars are the overlaps of the ground state
wave function described by Eq.(\ref{GROUND}) and the exciton wave
function described by Eq.(\ref{EXCITON}) with the corresponding
exact states at $L=0$ and $L=2$, respectively. The overlaps are near
unity. It should be noted, while Eq.(\ref{EXCITON}) also provides
angular momentum state at $L=3$, there is only one $L=3$ state in
the $(N_{\phi},N)=(9,6)$ finite system where the overlap is
trivially equal to $1$, and is therefore not shown in the figure.

\begin{figure*}[tbhp]
\label{Spectrum}
\includegraphics[width=18cm,height=14cm]{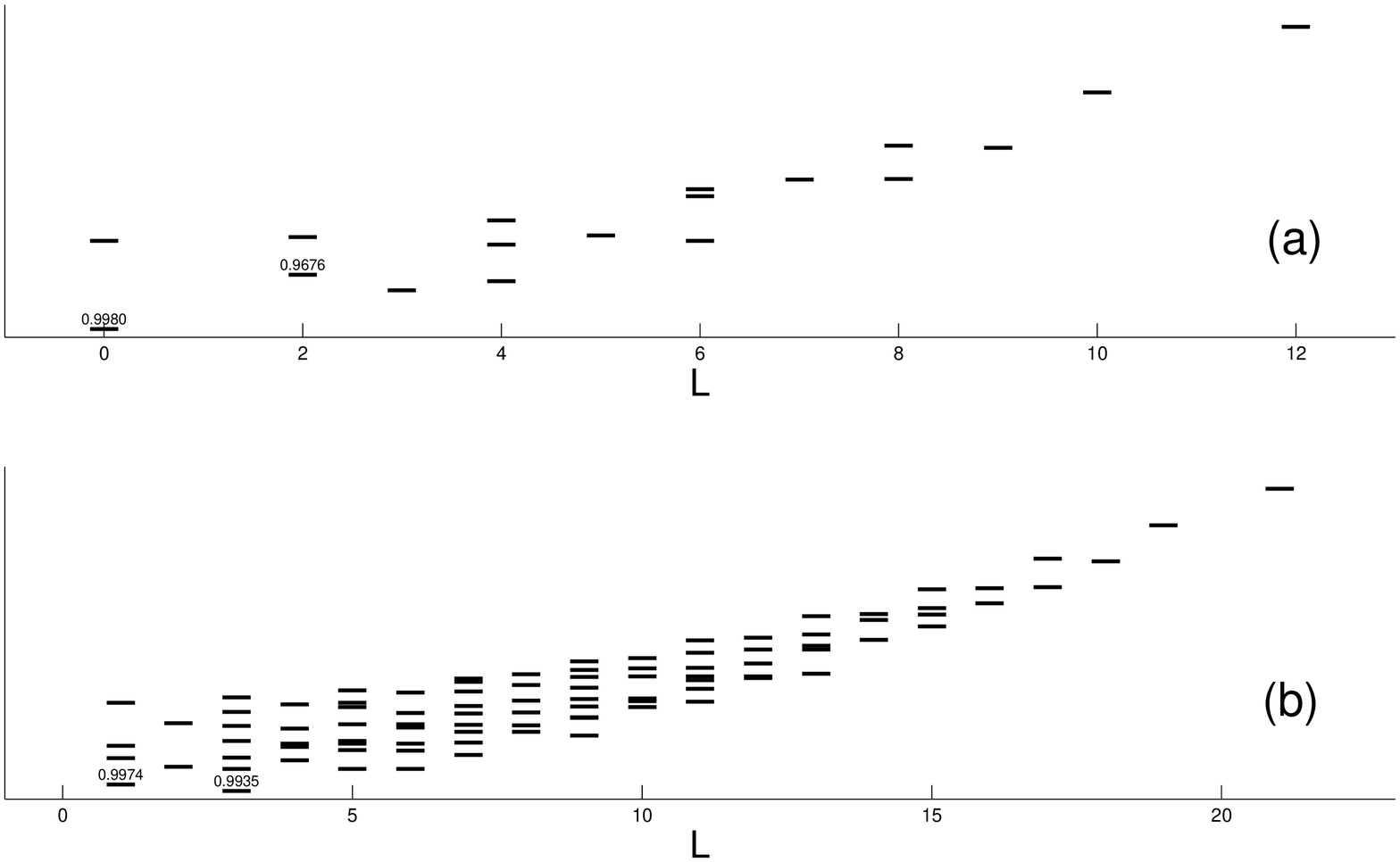}
\caption{(a) Energy spectrum of $(N_{\phi},N)=(9,6)$ finite system
in an arbitrary units versus angular momentum $L$. The numbers
($0.9980$ and $0.9676$) on top of the two energy bars are the
overlaps of the ground state wave function described by
Eq.(\ref{GROUND}) and the quasiexciton wave function described by
Eq.(\ref{EXCITON}) with the corresponding exact states at $L=0$ and
$L=2$, respectively. (b) Energy spectrum of $(N_{\phi},N)=(12,7)$
finite system. The numbers ($0.9974$ and $0.9935$) on top of the two
energy bars are the overlaps of the two fractionally charged QH wave
functions described by Eq.(\ref{TWOQH}) with the corresponding exact
states at $L=1$ and $L=3$, respectively.}
\end{figure*}

In Fig.1(b), we plot the energy spectrum for $(N_{\phi},N)=(12,7)$,
which corresponds to two $1/4$ fractionally charged QHs in the LLL.
In this case, Eq.(\ref{TWOQH}) describes two states in the $L_z=0$
sector with the angular momentum equal to $1$ and $3$. The overlaps
of them with the exact states are shown on top of the corresponding
energy bars. Again, exceedingly large overlaps are obtained.

We have also calculated overlaps for the finite systems shown in
Fig. 1(a) and 1(b) by varying the hardcore pseudopotential component
$V_1$ , while keeping other pseudopotential components at their LLL
Coulomb values. When $V_1$ is changed by a factor of $0.8$, $0.85$,
$0.9$, and $1.2$ from its LLL Coulomb value, the overlap in Fig.
1(a) at $L=0$ changes to $0.8961$, $0.9861$, $0.9948$, and $0.9996$,
respectively. The overlap in Fig. 1(a) at $L=2$ changes to $0.9856$,
$0.9878$, $0.9762$, and $0.9587$, respectively. The overlap in Fig.
1(b) at $L=1$ changes to $0.9280$, $0.9916$, $0.9976$, and $0.9938$,
respectively. The overlap in Fig. 1(b) at $L=3$ changes to $0.7970$,
$0.9833$, $0.9946$, and $0.9899$, respectively. In general, the
results show that the large overlaps have maintained until $V_1$
decreased below around $80\%$ of its corresponding Coulomb value,
manifesting the importance of the short range interaction to the
validity of the proposed wave functions. This shows a competing
requirement towards the formation of the incompressible FQHE state
where a certain degree of softening of the short range interactions
is needed. Therefore a delicate balance to meet both requirements is
important and that's probably one of the reasons why the FQHE at
half filling in the LLL has not yet been observed so far.

We have also calculated the overlaps in the SLL. They are $0.6730$
and $0.8608$ at $L=0$ and $L=2$ in Fig. 1(a), and $0.8074$ and
$0.6382$ at $L=1$ and $L=3$ in Fig.1(b). These rather moderate
overlaps are comparable to the performance of the MR states, making
them an alternative candidate to explain the observed $5/2$ FQHE.

Finally, we would like to briefly discuss situations when there is
one LQE left unpaired. This can happen when the total flux quanta
deviates from Eq.(\ref{NPHI_N}) by $1$ and $N$ is an even number. In
this case, the wave function is simply obtained by applying the LQE
operator Eq.(\ref{LQE_OP}) to Eq.(\ref{GROUND}). The unpaired
situation can also happen when Eq.(\ref{NPHI_N}) is satisfied but
$N$ is an odd number. In this case, there will be a LQE left
unpaired. There will also be two $1/4$ fractionally charged
quasiholes created. The corresponding wave functions can be obtained
by simply applying the LQE operator Eq.(\ref{LQE_OP}) to
Eq.(\ref{TWOQH}).

I would like to express my gratitude to Z. D. Wang for helpful
discussions, in particular for his hospitality during my visit in
Hong Kong University where this work was completed. I would also
like to thank D. N. Sheng for her help on finite size exact
diagonalizations. This work was supported by the GRF and CRF of the
RGC of Hong Kong.


\begin{thebibliography}{10}

\bibitem{Willett} R. Willett {\it et al}., Phys. Rev. Lett. {\bf59},1776 (1987).
\bibitem{MR} G. Moore and N. Read, Nucl. Phys. {\bf B360}, 362
(1991); N. Read and D. Green, Phys. Rev. {\bf B61}, 10267 (2000).
\bibitem{Greiter} M. Greiter, X.-G. Wen, and F. Wilczek, Phys. Rev. Lett. {\bf66},3205 (1991); Nucl. Phys. {\bf B374}, 567 (1992).
\bibitem{Morf} R. H. Morf, Phys. Rev. Lett. {\bf80}, 1505 (1998).
\bibitem{Peterson} M. R. Peterson, Th. Jolicoeur, and S. Das Sarma, Phys. Rev. Lett. {\bf101},016807 (2008); Phys. Rev. {\bf B78}, 155308 (2008).
\bibitem{Haldane} F. D. M. Haldane, Phys. Rev. Lett. {\bf51}, 605 (1983).
\bibitem{Sreejith} G. J. Sreejith {\it et al}., Phys. Rev. Lett. {\bf107}, 086806
(2011); G. J. Sreejith, A. W\'{o}js, and J. K. Jain, Phys. Rev.
Lett. {\bf107}, 136802 (2011).
\bibitem{Halperin} B. I. Halperin, P. A. Lee, and N. Read, Phys. Rev
{\bf B47}, 7312 (1993); E. Rezayi and N. Read, Phys. Rev. Lett.
{\bf72}, 900 (1994); J. Yang, Phys. Rev. {\bf B50}, 8028 (1994).
\bibitem{Laughlin} R. B. Laughlin, Phys. Rev. Lett. {\bf50}, 1395 (1983).
\bibitem{Levin} M. Levin, B. I. Halperin, and B. Rosenow, Phys. Rev. Lett. {\bf99}, 236806
(2007).
\bibitem{Lee} S.-S Lee, S. Ryu, C. Nayak, and M. P. A.
Fisher, Phys. Rev. Lett. {\bf99}, 236807 (2007).
\bibitem{Wang} Hao Wang, D. N. Sheng, and F. D. M. Haldane, Phys. Rev. {\bf B80},
241311(R),2009.
\bibitem{YangSuSu} J. Yang, Z. B. Su, W. P. Su, Mod. Phys.
Lett. {\bf B6}, 119 (1992).
\bibitem{Wen} X. G. Wen and A. Zee, Phys. Rev. {\bf B46}, 2290
(1992); X. G. Wen, Phys. Rev. Lett. {\bf70}, 355 (1993).
\bibitem{Wen2} X. G. Wen and A. Zee, Phys. Rev. Lett. {\bf69}, 953 (1992).
\bibitem{He} S. He, X. C. Xie, and F. C. Zhang, Phys. Rev. Lett. {\bf68}, 3460 (1992).
\bibitem{YangHierarchy} J. Yang, W. P. Su, Phys. Rev. Lett. {\bf70}, 1163
(1993); J. Yang, Phys. Rev. {\bf B49}, 16765 (1994).

\end{thebibliography}
\end{document}